# Simulating the Spread of Pandemics with Different Origins Considering International Traffic

Teruhiko Yoneyama
Multidisciplinary Science
Rensselaer Polytechnic Institute
Troy, New York, Unites States
yoneyt@rpi.edu

Mukkai S. Krishnamoothy
Computer Science
Rensselaer Polytechnic Institute
Troy, New York, United States
moorthy@cs.rpi.edu

*Abstract*—Pandemics have the potential to cause immense disruption and damage to communities and societies. In this paper, we propose a hybrid model to determine how the pandemic spread through the world. The model combines the SEIR-based model for local areas and the network model for global connection between countries. We simulate the potential pandemic with different origins and find how the difference of the origin of a pandemic influences the impact in the world. We investigate the travelers network which is derived from real data, and simulate 65 countries, and see how the pandemic spread through the world from different 14 countries as origins of pandemic. We compare the difference in terms of the impact in countries and the impact in the world. As a result, the impact in the world increases when pandemic originates from the United States, India, and China.

*Keywords-Simulation, Pandemic, SEIR, Social Network, International Traffic, Infectious Disease, Diffusion*

## I. INTRODUCTION

We simulate potential pandemics originating from different places. It is difficult to predict the origins of future pandemics. However, it may be possible to estimate how the spread progresses by tracing the number of travelers between countries. In this paper, at first we observe and analyze the travelers network which is derived from real data on the number of travelers between countries. Then we simulate a pandemic with several different origins using our model and figure out how the pandemic would progress from each of these different origins. We examine the impact of each pandemic in countries and in the world and compare the differences among the different origins.

## II. RELATED RESEARCH

Simulating the spreading of infectious disease has been studied in the past. We discuss the differences between this work and other related research. First, a lot of research about simulating disease spread focuses on a prevention/mitigation strategy by comparing the base simulation and an alternative simulation which considers their proposed strategy (e.g. [1][2][3][4][5][6]). In addition, most of existing research simulates with a generated situation which models the real world (e.g. [1][2][5][6][7][8][9]). On the other hand, we focus on simulating the pandemic using real data which reflects local condition and global connection. Although the simulation result could provide future hints that would help contain the spread of the disease, this paper does not directly propose a prevention strategy.

Second, much research considers the spread of infectious disease from either the local or global point of view (e.g. [3][5][6][8][10]). In addition, much research simulate using one of the equation based (e.g. SIR or SEIR differential equation model), agent based, or network based model (e.g. [8][11][12]). On the other hand, we simulate the pandemic from the global point of view considering local infection in each country. Also, we use a hybrid model which considers both the SEIR based model and network based model using the concept of agent based model.

Third, simulation parameters determine the path of spread. Some research values the basic reproduction number $R_0$ as an influential parameter (e.g. [3][13]). We don't determine $R_0$. This is based on the assumption that $R_0$ varies according to country. In our simulation, we consider setting the parameters to model a mild pandemic such that it doesn't spread so rapidly but spread worldwide.

## III. MODELING

Previous attempts to model spreading infectious diseases tended to fall into one of two categories. Equation-based models like the SEIR model is suitable for a large-scale spreading of diseases. These models use just a few parameters to reproduce the spreading phenomenon. However it is difficult to reflect detailed situation in countries which have different local infection conditions. Network or agent-based simulation models can theoretically reflect the detail of individual conditions. However, modeling large-scale global diseases is difficult as too many parameters are needed for simulation. Thus we propose a hybrid model. We make a simple model using a small number of parameters and make it capable of simulating a general pandemic.

We simulate using several countries. When we think of an infection in a country, there are three possibilities for new infection; (1) infection from foreign travelers, (2) infection from returning travelers, and (3) infection from local residents. Figure 1 illustrates this concept. We denote the infection-types (1) and (2) as the global infection and the infection-type (3) as the local infection.

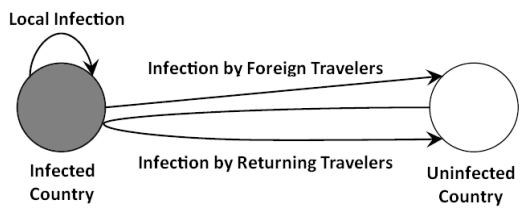

Figure 1: Three Patterns of Infection in a Country

We use the concept of SEIR model which considers four types of agents in each country; Susceptible, Exposed, Infectious, and Removed. Susceptible agents are infected by Infectious agents and become Exposed agents. Exposed agents are in an incubation period. After that period, Exposed agents become Infectious agents. Infectious agents infect Susceptible agents. Infectious agents become Removed agents after the infectious period. Removed agents are never infected again because they are now immune. Figure 2 illustrates this concept of SEIR model.

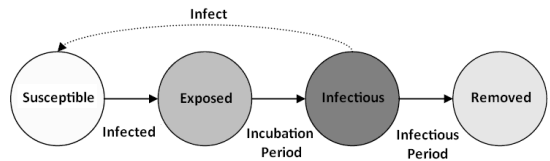

Figure 2: Concept of SEIR

At the beginning of the simulation, the number of Susceptible agents in each country is equal to the population of each country. Then we put an Infectious agent in the origin of the pandemic. The local infection spreads in the origin and the global infection also spreads from the origin to other countries through global traffic. When a country has at least one Infectious agent, that country has the potential for local infection. Figure 3 shows this concept.

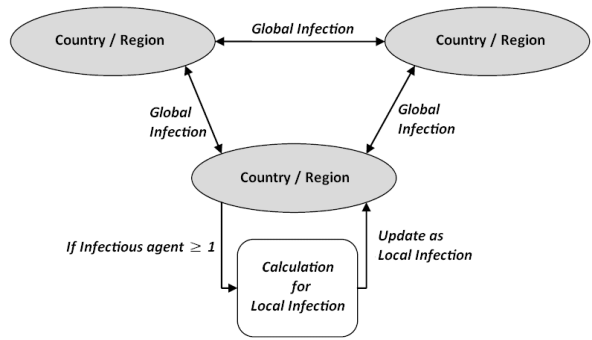

Figure 3: Concept of Simulation Task at One Cycle

The global infection is caused by traffic from infected country. Thus we refer to the number of inbound and outbound traffic. The number of new Exposed agents by the global infection in country $i$ at time $t$, $NEG_i(t)$, is calculated by the expression;

$$NEG_i(t) = I_j(t) \cdot T_{ij} \cdot P_G^*(t) \quad (1)$$

where $I_j(t)$ is the number of Infectious agents of country $j$ at time $t$. $T_{ij}$ is the total amount of both traffic from country $i$ to $j$ and from $j$ to $i$. $P_G^*(t)$ is the global infection probability at time $t$ and is calculated by the expression;

$$P_G^*(t) = P_G - (D_G \cdot t) \quad (2)$$

where $P_G$ is the basic global infection probability between countries. $D_G$ is a "deductor" for the global infection. $t$ is time (simulation cycle). $P_G$ and $D_G$ are constants and are uniformly used for every country. Thus the global infection probability $P_G^*(t)$, decreases along the simulation cycle. We assume that, in the real world, the global infection occurs with high probability in early pandemic due to the lack of awareness of the disease. As the disease spreads, people take preventive measures against the infection and the pandemic decreases. We apply this concept in the simulation. The number of Exposed agents in country $i$ at time $t$, $E_i(t)$, is updated by adding $NEG_i(t)$ to $E_i(t)$ at each simulation cycle.

We assume that the local infection probability depends on the population density of a country. Thus if the country is dense, people are more likely to be infected. The basic local infection probability of country $i$, $P_{Li}$ is given by the expression;

$$P_{Li} = Density_i \cdot C_1 + C_2 \quad (3)$$

where $Density_i$ is population density of country $i$, obtained by real data. Thus $Density_i$ differs in country. $C_1$ and $C_2$ are constants and are used for simulation in every country.

We assume that the number of new Exposed cases of a country by the local infection depends on the number of Susceptible agents and the number of Infectious agents at that time. Thus the number of new Exposed agents by the local infection in country $i$ at time $t$, $NEL_i(t)$, is calculated by the expression;

$$NEL_i(t) = S_i(t) \cdot I_i(t) \cdot P_{Li}^*(t) \quad (4)$$

where $S_i(t)$ us the number of Susceptible agents of country $i$ at time $t$. $I_i(t)$ is the number of Infectious agents of country $i$ at time $t$. $P_{Li}^*(t)$ is the local infection probability at time $t$ and is calculated by the expression;

$$P_{Li}^*(t) = P_{Li} - (D_L \cdot t) \quad (5)$$

where $P_{Li}$ is the basic local infection probability of country $i$ which is obtained by equation (3). $D_L$ is a "deductor" for the local infection and is a constant which is used for every country. $t$ is time (simulation cycle). Similar to the global infection, the local infection probability $P_{Li}^*(t)$ decreases as the simulation cycle increases. This reflects people's awareness. The number of Exposed agents in country $i$ at time $t$, $E_i(t)$, is updated by adding $NEL_i(t)$ to $E_i(t)$ at each simulation cycle.

Table 1 summarizes parameters in the simulation. We have eight controllable parameters which are denoted as constants in Table 1. These parameters are used for every country uniformly. Other parameters are derived from real data and depend on country.

Table 1: Parameters in Simulation

| Parameter | Description | Attribution | |
|---|---|---|---|
| | | (a) Global or (b) Local | (1) Constant or (2) Depend on Country |
| $P_G$ | Global Infection Probability | (G) | (1) |
| $P_{Li}$ | Local Infection Probability of County $i$ | (L) | (2) |
| $D_G$ | Deductor for Global Infection Probability | (G) | (1) |
| $D_L$ | Deductor for Local Infection Probability | (L) | (1) |
| $C_1$ | Constant for Local Infection Probability | (L) | (1) |
| $C_2$ | Constant for Local Infection Probability | (L) | (1) |
| *Incubation_Period* | Incubation Period | (G) and (L) | (1) |
| *Infectious_Period* | Infectious Period | (G) and (L) | (1) |
| *Run_Cycle* | Run Cycle of Simulation | (G) and (L) | (1) |
| *Density$_i$* | Actual Population Density of Country $i$ | (L) | (2) |
| *Population$_i$* | Actual Population of Country $i$ | (L) | (2) |
| $T_{ij}$ | Amount of Traffic between Country $i$ and $j$ | (G) | (2) |

IV. TRAVELERS NETWORK

First, we select some origins for the simulation. In order to select countries from different geographical areas, we choose 5 of the continents: Africa, Americas, Asia, Europe, and

Oceania. From these 5 regions, we choose 2 to 3 countries based on highest GDP countries referring to IMF's estimations [14] and choose 14 countries; Australia, Brazil, Canada, China, Egypt, France, Germany, India, Japan, New Zealand, Nigeria, South Africa, United Kingdom, and the United States. Then we pick the Top 5 countries which are most strongly related with these 14 countries in terms of the number of travelers, referring to [15][16]. For example, Top 5 related countries based on the travelers with the United States are Mexico, Canada, United Kingdom, France, and Japan. Then we find 41 countries including first 14 countries. Next we repeatedly examine Top 5 related countries with these 41 countries. As an example, Mexico's related countries are the United States, Canada, France, Spain, and United Kingdom. Thus we find 2 neighborhoods from given 14 countries. As a result, we find 65 countries in total.

We investigate the travelers network among 65 countries. Figure 4 shows the travelers network on a geographical map. In this map, countries are connected with a line if there are some travelers. The thickness of the line differs based on the number of travelers between countries.

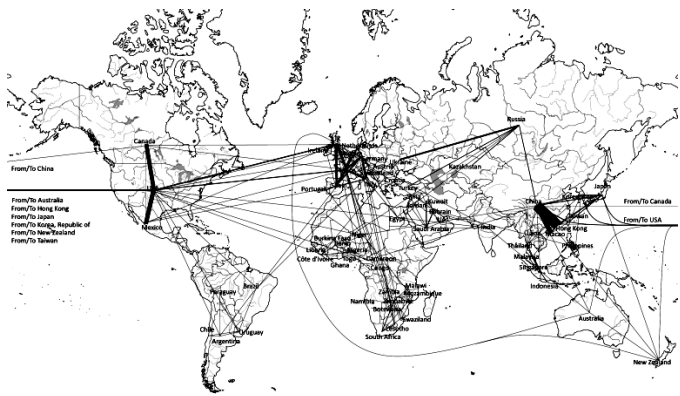

Figure 4: Travelers Network among 65 Countries

There are some remarkable thick links. The strongest relationship is between China and Hong Kong. In fact, the number of travelers between China and Hong Kong is 87,041,608 in 2007. The relationship between China and Macao is also very high: 38,053,307 [15][16]. Next, the relationships between the United States and its adjacent countries are strong. The total numbers of travelers with the United States are 33,785,629 in Mexico, and 31,132,366 in Canada. Also, the relationship between Canada and Mexico is not week; 1,199,916 [15][16]. The relationships between European countries are also strong. In particular, there are many travelers among United Kingdom, France, Germany, Italy, and Spain. Other European countries also have some relationship with these five countries. Russia has a relatively strong relationship with the Ukraine. Thus, when one of these countries is infected, the neighboring countries and areas are infected quickly.

Next, we observe the network in each region. In Africa, there are two dense networks. One is Nigeria and its neighboring countries. The other is South Africa and its neighboring countries. Although each of these networks is dense within the network, these two networks are not directly connected. Egypt is not included in both networks. Egypt is closer to Europe, Saudi Arabia, and Russia than other African countries. In Americas, the United States, Canada, and Mexico have strong relationship with each other. North America and South America are connected by the United States. There are connections between the United States and Brazil, Argentina, and Uruguay, and it is expected that the infection spreads between North and South America through these countries. In Asia, countries in Far East and Southeast are thickly connected to China. Excluding China, there are three different networks in Asia. The first is the Far Eastern network which includes Japan, Republic of Korea, Taiwan, Hong Kong, and so on. The second is the Malayan network which includes Thailand, Malaysia, Singapore, and Indonesia. The third is the Middle Eastern network which is centering Saudi Arabia. Bahrain is somewhat more connected with United Kingdom. China thickly connects the Far Eastern network and the Malayan network. India connects the Malayan network and the Middle Eastern network. India is also connected with the United States and Europe. In Europe, countries are thickly connected centering on United Kingdom, France, Germany, Italy, and Spain. Russia has strong relationship with the old USSR countries, and has some relationships with China and Turkey. In Oceania, Australia and New Zealand has relationship and they also have some strong connections with Asia, Americas, and Europe,

Next we find out some important intercontinental links. The United States has strong relationships with United Kingdom and France. Canada has also relatively strong relationships with these countries. Thus when the pandemic originates from North America, it is expected that it quickly spreads to Europe through United Kingdom and France. Similarly, the opposite direction is expected (links are bidirectional). The United States has a strong connection with Far Eastern countries and Oceania. In addition, the United States has a connection with India and Canada, which have a connection with China. In the simulation, these connections play an important role for the spread of a pandemic. The United States has some relations with South American countries and African countries. European region is widely connected with many countries in other regions. For example, Germany and France have some relations with African countries. In particular, United Kingdom has many connections with African countries, Australia, New Zealand, and India, in addition to North American countries. United Kingdom and Germany are connected with Egypt, and Egypt is connected with Russia. This route is possibly influential for Russia. Similarly, the connection between China and Australia is also possibly influential.

## V. SIMULATION AND RESULTS

We simulate 65 countries. For the global infection, we use the travelers network which is investigated in the previous section. For the local infection, we use the actual population and population density in each country simulated referring to [17]. We set the simulation parameters to model a mild pandemic such that it spreads worldwide but does not spread so rapidly. We set the simulation cycle as 364, which is long enough to simulate possible converges of the pandemic.

In the simulation with 14 different origins, we observed some general characteristics regarding the infection route throughout the world. At first, some regions have some countries which play the role of an "entrance" or an "exit". For example, in Europe, United Kingdom and France are often infected at first among European countries and spreads to other European countries. Moreover, the pandemic tends to spread to other regions through these countries. In the European pandemic, United Kingdom contributes to the worldwide spread because of its large number of links. United Kingdom can be an exit to Africa since pandemics which occur outside Africa often spread to Africa through United Kingdom. Typically, the pandemic spreads to Nigeria through United Kingdom. Although the entrance or exit to Europe is the United States, United Kingdom, and France, sometimes Italy and Spain can also serve entrance/exit. For example, when the pandemic originates from Brazil, these two countries are infected earlier than other European countries. China contributes to spread of the pandemic to other Asian countries. The United States is the entrance to South America: when a pandemic occurs outside South America, South American countries are often infected through the United States. Australia can be an entrance/exit of Oceania. Although Russia is large country, there are only two infection routes when we simulate with 65 countries and given 14 origins. One is from Egypt and another is from China. Thus these countries serve as the entrance of Russia.

Another observation from the simulation with 14 different origins is that there are some countries which are infected by a particular country regardless of the origin of the pandemic. For example, Philippines are always infected from Republic of Korea. Some African countries are usually infected by Nigeria or South Africa. Malaysia is often infected through Thailand or Singapore. Germany tends to be infected through France. These tendencies are because of the limitations of the connections. Some countries have only one link and can thus only be infected through that connection. In this experiment, we simulate 65 countries considering Top 5 related countries in 41 countries. These tendencies may change when we take into account more countries or relations.

Next, we compare the impact in each country with varying pandemic origins. The absolute number of cases in a country can be obtained from the simulation result. However that is not important since that varies in the set of parameters. The given set of parameters is one example for the comparison. The differences in the impact of the pandemic to different countries are more important than the raw number of cases. Thus we look at the ratio of the number of cases in a country to that in the origin. Our results can be useful when comparing different situations. For example, in this paper, we discover that certain countries have more influence than others. In the following section, we focus on some countries which have most significant impact with different situations, but not specifically on the number of cases in each situation.

Table 2 shows Top 5 countries which have most significant impact with 14 different origins. Each country has a value which shows the ratio of the number of cases in that country to that in the origin. For example, when Brazil is an origin of the pandemic, the United States is the most significantly impacted country in the world and the number of cases is 1.58 times larger than that of Brazil, while Japan is thirdly impacted and its number of cases is 67% of Brazil. When the origin is New Zealand, the infection does not spread to other countries in our simulation.

In the simulation result, every country is most significantly impacted when it is the origin of the pandemic. For example, the Unites States is most significantly impacted when the pandemic originates in the United States. However, it is not necessary that the country of origin of a pandemic is the most seriously affected. For example, when the origin is the United States, the country which has the most significant impact is China.

Table 2: Top 5 Impacted Countries with 14 Different Origins

| Origin | Top 5 Impacted Countries (Value: Ratio of Number of Cases to That in Origin) | | | | |
|---|---|---|---|---|---|
| | 1 | 2 | 3 | 4 | 5 |
| Australia | China 17,988.68 | UK 827.09 | US 114.28 | Singapore 61.96 | Australia 1.00 |
| Brazil | US 1.58 | Brazil 1.00 | Japan 0.67 | Germany 0.43 | France 0.34 |
| Canada | China 2,583.78 | US 588.63 | Japan 249.09 | Mexico 212.47 | France 124.55 |
| China | China 1.00 | India 0.85 | Indonesia 0.18 | Japan 0.10 | Philippines 0.07 |
| Egypt | Germany 1.03 | Egypt 1.00 | France 0.79 | UK 0.76 | Italy 0.72 |
| France | US 4.73 | Japan 2.00 | Mexico 1.71 | Germany 1.29 | Egypt 1.26 |
| Germany | Germany 1.00 | Egypt 0.97 | France 0.77 | UK 0.74 | Italy 0.71 |
| India | China 1.17 | India 1.00 | US 0.27 | Indonesia 0.21 | Japan 0.11 |
| Japan | China 10.37 | US 2.36 | Japan 1.00 | Philippines 0.71 | Thailand 0.51 |
| New Zealand | New Zealand 1.00 | NA | NA | NA | NA |
| Nigeria | Nigeria 1.00 | UK 0.45 | Benin 0.06 | Cameroon < 0.001 | Niger < 0.001 |
| South Africa | South Africa 1.00 | Lesotho 0.002 | Zimbabwe < 0.001 | Mozambique < 0.001 | Swaziland < 0.001 |
| UK | India 18.59 | US 4.95 | Nigeria 2.22 | Japan 2.10 | Mexico 1.79 |
| US | China 4.39 | India 3.75 | US 1.00 | Japan 0.42 | Mexico 0.36 |

Next, we estimate and compare the number of cumulative cases in the world with different origins. Figure 5 shows the comparison of the estimated total cases in the world with 14 different origins. X-axis shows the origins. Y-axis shows the ratio to the total number of infected cases in the world in case the United States is the origin. For example, suppose that the number of cases is 100 when the origin is the United States, and that is 90 when the origin is country X. Then, the ratio is 90%. Thus we compare the impact with different origins based on the United States case since the world has the largest infected cases when the United States is the origin in the simulation result. We compare the difference of origins by ratio since the number of cases is not meaningful since the number of cases varies depending on the set of parameters. When calculate the total number of cases in the world as the total population in the world is 6.67 billion, while the total population in simulated 65 countries is about 5 billion, referring to [17].

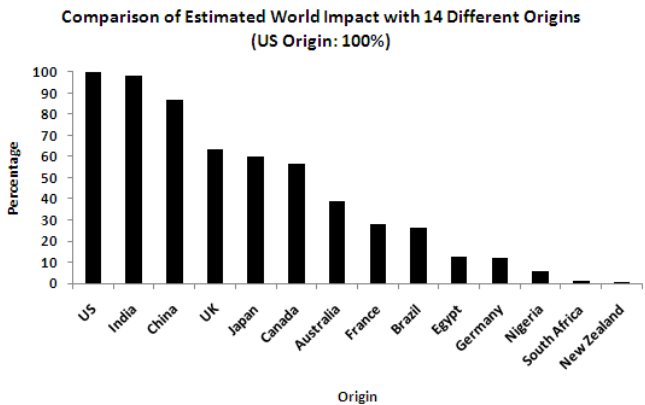

Figure 5: Comparison of Estimated World Impact with 14 Different Origins assuming US Origin is 100%

Comparing between different origins, we figure out that the world is most impacted when the pandemic originates in the United States. When the United States is the origin of the pandemic, it is expected that the United States has a large number of local infections because of its large population. In addition, the United States has many related countries to infect quickly. In order of impact to the world, the next few countries are India and China. It is thought that this is largely influenced by the local infection in China and India, with their huge population.

The fourth influential country is United Kingdom. This is because United Kingdom has many related countries to spread the pandemic. Japan and Canada is ranked fifth and sixth in this figure. These countries strongly related with China and the United States. Thus when the pandemic occurs in these countries, it quickly spreads to China and the United States at first, and then spreads worldwide from these countries. Consequently the number of cases in the world increases. In fact, looking at Table 2, we see China and the United States are the most impacted countries when the origin is Canada and/or Japan. Thus when the pandemic originates from these countries, it quickly spreads to all over the world and the total cases in the world increases.

The world has the least impact when the origin is New Zealand in this experiment. It is thought that New Zealand has less potential to have the local infection due to its small population. Thus an infection hardly spreads out from New Zealand to other countries in our simulation. When we use higher value for the parameter of the infection probability for local and global, it is expected that the pandemic spreads to the world through Australia. In addition, the world has less impact when the origin is South Africa in our experiment. This is due to its less relationship with other regions. In our simulation, the epidemic originates from South Africa spreads to only its neighboring countries and epidemic does not spread worldwide. When we use higher value of the global infection probability, it is expected that the pandemic spreads to the world through United Kingdom and the total cases in the world increases.

VI. CONCLUSIONS

In this paper, we simulated pandemics originating 14 different countries. We referred to the Top 5 related countries in terms of the number of travelers in real data. We found totally 65 countries which were 2 neighborhoods from given 14 countries. Thus we simulated these 65 countries.

We analyzed the travelers network among 65 countries and found some strong relationships between particular countries. For example, China, Hong Kong, and Macao had a strong relationship to each other, as do the United States, Canada, and Mexico. European countries were also strongly tied, especially among United Kingdom, France, Germany, Italy, and Spain. The United States had a strong relationship with United Kingdom, France, India, and some Far Eastern countries and these links contributed to the spread of the pandemic. Russia also had strong relationships with China and some old USSR countries.

Although the infection route varied in the different origins, we found out that there were some general tendencies and some important countries for the spread. For example, United Kingdom and France were often infected first among European countries and spread the pandemic to other

European countries. United Kingdom also contributed to spread the pandemic to other European countries. United Kingdom also contributed to spread the infection to African countries. In many cases, the United States spread the infection to South America. In Asia, China often played the role of the entrance/exit of the pandemic.

The impact of the pandemic differed with the origin. We compared the impact of the pandemics in each country with 14 different origins. The impact in a country much depended on the origin. Then we found that each country had the most significant impact when it was the origin, but it was possible that other countries had more impact than the origin had. Next we compared the impact of each of the pandemics to the world by estimating the total number of cases in the world for each origin case. We discovered that world had the most significant impact when the pandemic originated in the United States in our simulation. It is thought that this is because the United States has many related countries. The estimated number of cumulative cases in the world was larger, when the origin was the United States, India, China, United Kingdom, Japan, and Canada. These countries have, a large population, high population density, and/or, perhaps most importantly, many related countries. When the pandemic originated from these countries, it quickly spread to all over the world and the total number of cases in the world was large. The impact of the pandemic is influenced by the characteristics of the specific infectious disease such as the infection rate and mortality rate. We expect that the location of the origin of the pandemic also strongly influences the impact to the world since the total number of cases in the world varies based on different origins of the pandemic, as our simulation results showed. Thus the investigation of the infection routes considering the travelers network and the number of travelers between countries is important to understand how a pandemic spreads through the world.